\documentclass[12pt]{iopart}

\usepackage{graphicx}
\usepackage{color}
\usepackage{iopams}  
\usepackage{exscale}
\usepackage{makeidx,shortvrb,latexsym}
\usepackage{epstopdf}
\usepackage{tabularx, booktabs, multirow}
\usepackage[]{hyperref}

\usepackage{graphics}

\def\pmb#1{\setbox0=\hbox{#1} 
\kern-.020em\copy0\kern-\wd0 
\kern-.04em\copy0\kern-\wd0 
\kern-.020em\raise.0383em\box0}

\begin{document}

\title[Dislocation/precipitate interactions in Mg-Al alloys]{Basal dislocation/precipitate interactions in Mg-Al alloys: an atomistic investigation}

\author{Gustavo Esteban-Manzanares$^{1,2}$, Anxin Ma$^1$, Ioannis Papadimitriou$^1$, Enrique Mart{\'\i}nez$^3$, Javier LLorca$^{1, 2}$ \\}
\vspace{10pt}
\address{$^1$IMDEA Materials Institute, C/ Eric Kandel 2, 28906 - Getafe, Madrid, Spain\\\ \\
$^2$Department of Materials Science, Polytechnic University of
  Madrid/Universidad Polit�\'ecnica de Madrid, E. T. S. de Ingenieros de Caminos. 28040 - Madrid, Spain. \\\ \\
  $^3$ Theoretical Division T1, Los Alamos National Laboratory, Los Alamos 87545 NM, USA.}

\vspace{10pt}

\begin{indented}
\item[]June 2019
\end{indented}

\begin{abstract}
The interaction between edge basal dislocations and  $\beta$-Mg$_{17}$Al$_{12}$ precipitates was studied using atomistic simulations. A strategy was developed to insert a lozenge-shaped Mg$_{17}$Al$_{12}$ precipitate with Burgers orientation relationship within the Mg  matrix in an atomistic model ensuring that the matrix/precipitate interfaces were close to minimum energy configurations. It was found that the dislocation bypassed the precipitate by the formation of an Orowan loop, that entered the precipitate. Within the precipitate, the dislocation was not able to progress further until more dislocations overcome the precipitate and push the initial loop to shear the precipitate along the (110) plane, parallel to the basal plane of Mg. This process was eventually repeated as more dislocations overcome the precipitate and this mechanism of dislocation/precipitate interaction was in agreement with experimental observations. Moreover, the initial resolved shear stress to bypass the precipitate was in agreement with the predictions of the Bacon-Kocks-Scattergood model. 

\end{abstract}

\vspace{2pc}


\submitto{Modelling and Simulation in Materials Science and Engineering}

\maketitle

\section{Introduction}

Dispersion of nm-sized precipitates is known to be a very effective strategy to increase the flow strength of metallic alloys \cite{KN63}. Al-Cu and Ni-based superalloys are excellent examples of this behavior but similar success has not been achieved in other metallic materials such as Mg alloys \cite{N12, GZM05, BGV04, MOK09, JMO10}. Mg  has a HCP lattice and the deformation mechanisms are very different from those found in FCC alloys \cite{C68, GWE98, SGC12}. In particular, plastic deformation of Mg and Mg alloys takes place by dislocation glide in three different slip systems (basal, prismatic and pyramidal slip) as well as by tension twinning as opposed to FCC metals in which slip occurs in \{111\} planes along the $<$110$>$ directions. The large differences in the critical resolved shear stresses  (CRSS)   to activate plastic slip in the different systems in Mg as well as the polarity of twinning (which only takes place when the $c$ axis of the HCP lattice is extended), lead to the plastic anisotropy of Mg alloys, that has very negative effects on the ductility \cite{HB10, HLD14}.

Basal slip is the softest slip system in Mg, and increasing the  CRSS   for dislocation slip in this system is necessary to improve the yield strength of Mg alloys and to reduce the plastic anisotropy \cite{HB10}. Precipitation hardening has been used to this purpose but the contribution of precipitates to strengthen basal slip has been limited, particularly in Mg-Al alloys  \cite{HNG05, HHP14, RLK16, HRP17}. Continuum models, based on the Orowan mechanism, have been applied to ascertain the influence of the precipitate size, shape, orientation and spatial distribution on the  CRSS   on the different slip systems in Mg \cite{N03, SGC12, HRP17}. Although these models were able to rationalize some of the experimental trends, it should be noticed that the Orowan model assumes a very simplistic approach for the dislocation/precipitate interaction: the precipitates are rigid obstacles overcome by dislocations by the formation of an Orowan loop. Nevertheless, recent micropillar compression tests in Mg - 5 wt.\% Zn alloys \cite{WS15, AL19} have shown evidence of precipitate shearing basal dislocations and similar results were found in transmission electron microscopy observations of dislocation/precipitate interactions in Mg-Al alloys \cite{VKM17, CCP19}. Thus, simulations at smaller length scales (either using dislocation dynamics \cite{XSC04, LGL16, HRU17, SEP18} or molecular mechanics \cite{HSC00, SW10, BTM11, LLH14, LGL16, EMS19}) are required to get a better understanding of the physics and to develop more accurate models. 

While molecular mechanics is a very attractive approach to understand the details of the mechanisms of dislocation/precipitate interaction, the validity of the simulations depends on a number of factors. First, it is important to ensure that the interatomic potential is able to reproduce accurately, also from a quantitative viewpoint, the main physical parameters that control the dislocation/precipitate interaction. They  include the Peierls stress and the stacking fault energy, the elastic constants of matrix and precipitate as well as the interface energies for the different matrix/precipitate interfaces along the dislocation path. Second, the details of the matrix/precipitate interface (that also are related to the precipitate shape) are very important, in so far they can modify the progress of the dislocation. The matrix/precipitate interface is well defined from the atomistic viewpoint in the case of coherent precipitates but not in the case of incoherent ones and a consistent strategy should be developed to build the interface in this case. Finally, molecular statics (that do not take into account thermal vibrations) and molecular dynamics (that are carried out to very high strain rates) simulations may or may not lead to different results and these differences should be analyzed. All these factors have to be taken into account because the credibility of the atomistic simulations is otherwise limited.

In this investigation, molecular statics and dynamics simulations were carried out using a new interatomic potential \cite{DD18} to assess the interaction between basal dislocations and $\beta$-Mg$_{17}$Al$_{12}$ intermetallic precipitates in Mg-Al alloys, which are well-known among Mg alloys for their excellent castability and corrosion resistance \cite{BD08, N12}. The validity of the new interatomic potential to simulate dislocation/precipitate interactions is first demonstrated and compared with other potentials available in the literature. Afterwards, the strategy to introduce precipitates with stable interfaces based on the experimental information is presented and applied to $\beta$-Mg$_{17}$Al$_{12}$ precipitates in a Mg matrix. Then, molecular statics and dynamics simulations are carried out to ascertain the mechanisms of dislocation/precipitate interaction as a function of the position of the slip plane, temperature and distance between precipitates.

\section{Background on precipitate-strengthened Mg-Al alloys}

\begin{figure}
	\centering
		\includegraphics[width=\textwidth]{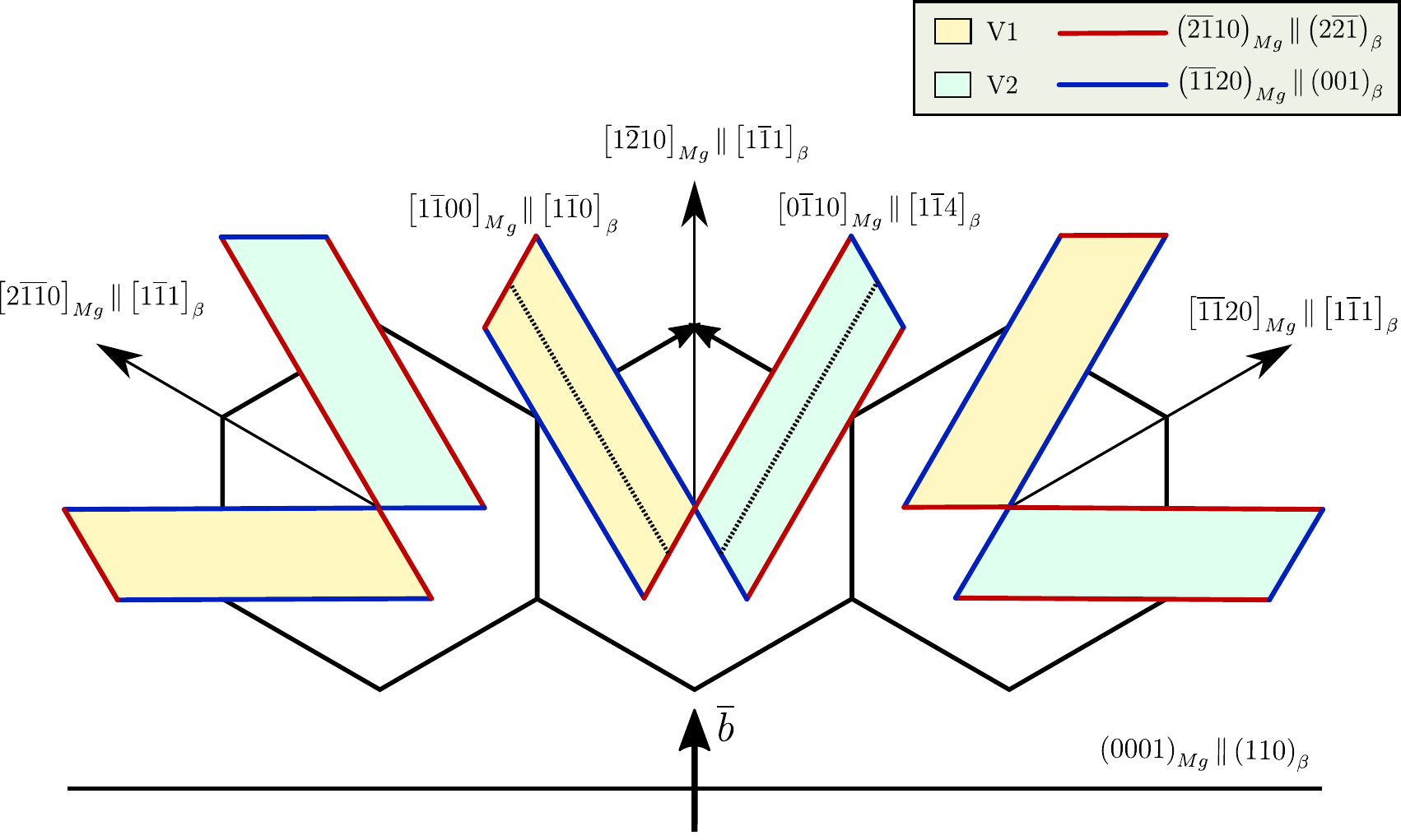}
	\caption{Schematic of the orientation and interface planes of the different variants (V1 and V2) of a lozenge-shape $\beta$ precipitate in the Burgers OR. The edge basal dislocation is shown as a horizontal black line.}
	\label{Var}
\end{figure}

Precipitation of Mg$_{17}$Al$_{12}$ ($\beta$) intermetallic occurs in Mg-Al alloys when the cooling rate of the casting is sufficiently slow or by means of aging treatments after casting \cite{N12}. The $\beta$ phase has a complex BCC structure within the space group $I-43m$ and with a lattice parameter of 1.056 nm. The unit cell includes 34 Mg atoms and 24 Al atoms \cite{BM08}. The generalized stacking fault energies ($\gamma$ surfaces) corresponding to the potential slip systems were recently determined using first principles methods and compared with the Griffith surface fracture energy \cite{xiao2013}. It was concluded that  high shear stresses were necessary to promote dislocation slip in this intermetallic  and that its behavior was brittle in the presence of a crack because it was more likely to show crack propagation than to emit dislocations from the crack tip, according to Rice's criterion \cite{R92},

These precipitates grow in the matrix along preferred orientations. Different orientation relationships (OR) and precipitate shapes has been reported in the literature \cite{crawley1974I, crawley1974II, duly1994I, duly1994II, duly1995, shepeleva2001, zhang2003} but the most typical one is the Burgers OR. $\beta$ precipitates in Burgers OR have a lozenge-shape \cite{C00, crawley1974I} and, according to the X-ray diffraction pattern,  the growth habit plane is given by  $(0001)_{Mg} \| (110)_{\beta}$  with a coincident direction $\left[1\overline{2}10\right]_{Mg} \| \left[1\overline{1}1\right]_{\beta}$. This direction stands as a symmetry axis of two variants that grow in $[1\overline{1}00]_{Mg} \| [1\overline{1}0]_{\beta}$ (V1) and $[0\overline{1}10]_{Mg} \| [1\overline{1}4]_{\beta}$ (V2) directions along the same habit basal plane (Fig. \ref{Var}). Thus, the lateral interface planes of these variants (perpendicular to the basal plane of Mg) are $(\overline{1}\overline{1}20)_{Mg} \| (001)_{\beta}$ and $(\overline{2}\overline{1}10)_{Mg} \| (2\overline{2}\overline{1})_{\beta}$, as shown in Fig. \ref{Var}. It should be indicated that the growth directions of V1 and V2 present a misorientation of  5$^{\circ}$ with respect to the $<$1100$>$ direction of the Mg lattice. 12 different variants of the $\beta$ precipitate with Burgers OR can be distinguished, combining the different possible interfaces. From the viewpoint of the interaction of an edge basal dislocation with the precipitate, there are 6 independent orientations in which the growth directions of V1 and V2 precipitates form angles of $0^\circ$, $60^\circ$ and $-60^\circ$ with Burgers vector of the dislocation (Fig. \ref{Var}).   

Although noticeable, the age hardening response of Mg-Al alloys is not as good as it could be desired. It has been argued that the limited precipitation hardening of Mg-Al alloys occurs because most of the  precipitates  exhibit Burgers OR and lay parallel to the basal plane, and this orientation is the least efficient to block basal slip \cite{HNG05}. It  was also suggested that the precipitate distribution is relative coarse because of the  high  diffusion  rate  of  Al  atoms  in  the  Mg  matrix \cite{wang2016}. Finally, based on the Orowan model, it was proposed that rod-shaped precipitates that grow perpendicular to the basal plane with Crawley OR  ($\left\{0001\right\}_{Mg} \|\left\{111\right\}_{\beta}$; $\left\langle1210\right\rangle_{Mg} \| \left\langle112\right\rangle_{\beta}$) are more efficient because they intersect a larger number of basal planes for a given precipitate volume fraction \cite{C00, HNG05, N03, robson}. Nevertheless, previous atomistic simulations of the interaction of basal and prismatic edge dislocations with $\beta$ precipitates did not show any evidence of the formation of Orowan loops \cite{LLH13, LLH14}. Thus, the  details of the dislocation/precipitate interaction in these alloys as well as the strengthening provided  are not well known.

\section{Methods and Model development}

\subsection{Development of the atomistic matrix/precipitate model}\label{AM}
 
\begin{figure}
	\centering
		\includegraphics[width=0.6\textwidth]{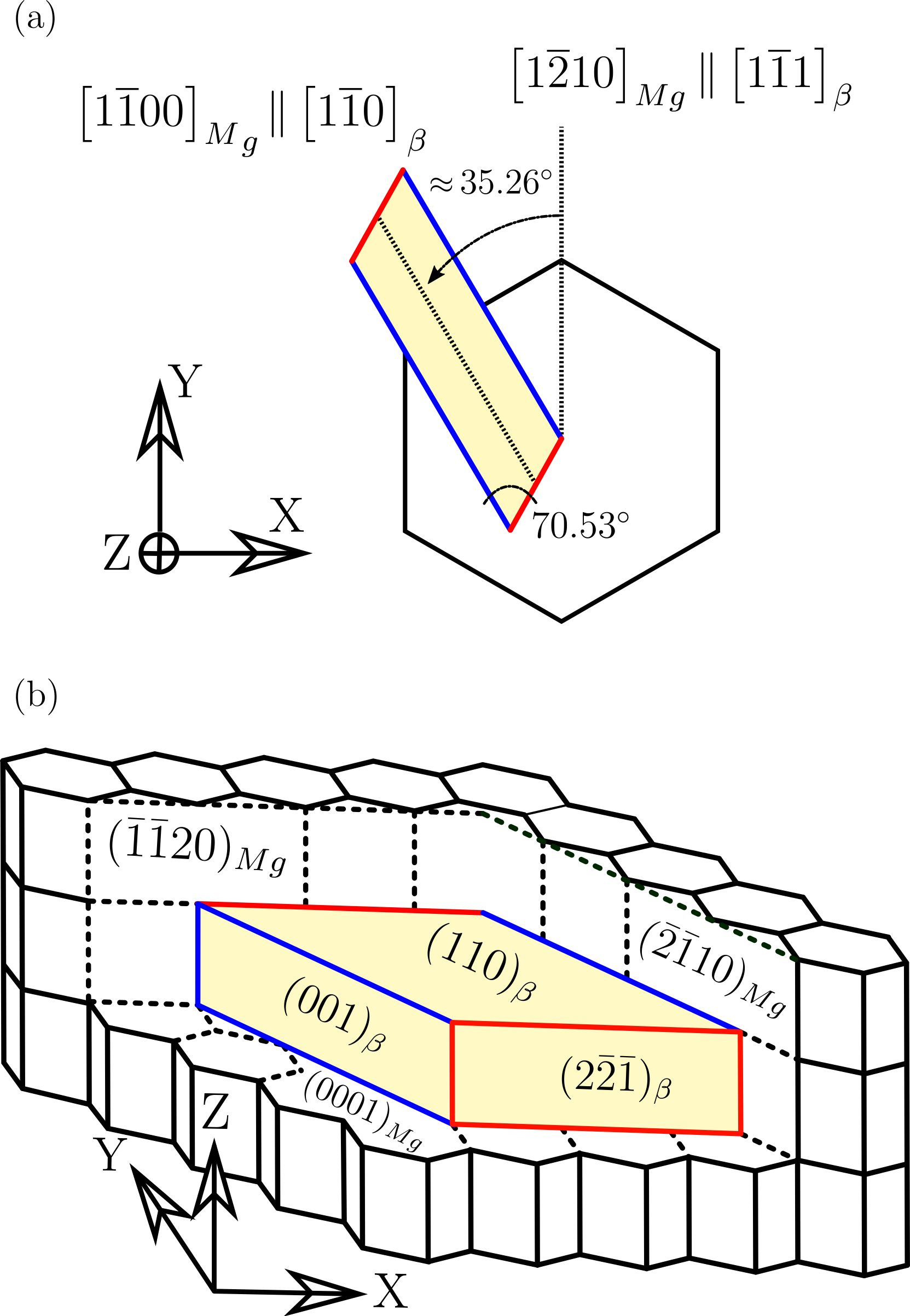}
	\caption{(a) Orientation of the $\beta$-Mg$_{17}$Al$_{12}$ lozenge-shaped precipitate with respect to Mg matrix in the atomistic model. The $Z$ axis is perpendicular to the basal plane of the Mg lattice. (b) 3-D representation of the $\beta$ precipitate embedded in the HCP Mg matrix. The coincident crystallographic planes are denoted on both the $\beta$ precipitate (yellow) and the Mg matrix (white).}
	\label{Rot}
\end{figure}

A sequential strategy was used to insert a lozenge-shaped $\beta$ precipitate with Burgers OR within the Mg matrix (Fig. \ref{Var}) ensuring that the semicoherent interfaces have minimum energy. This step is critical to assess the interaction between the dislocations and the precipitate because the propagation of the dislocation along the interface will depend on the interface features. The dimensions of the precipitate were $12.4\times 3.14 \times 2.8$ nm$^3$, with an aspect ratio 4:1 parallel to the basal plane in agreement with the experimental observations \cite{LLH09}. The precipitate was inserted in a small periodic Mg domain of dimensions $19 \times 17 \times 8$ nm$^3$ along the $X$, $Y$ and $Z$ axes, respectively, which were oriented parallel to $[10\bar{1}0]$, $[1\bar{2}10]$ and $[0001]$ directions of the Mg lattice. The $[1\overline{1}0]$ direction (long edge) of the precipitate was aligned with the $[1\overline{1}00]$ direction of the matrix, as depicted in Fig. \ref{Rot}. 

\begin{figure}[h!]
	\centering
		\includegraphics[scale=0.8]{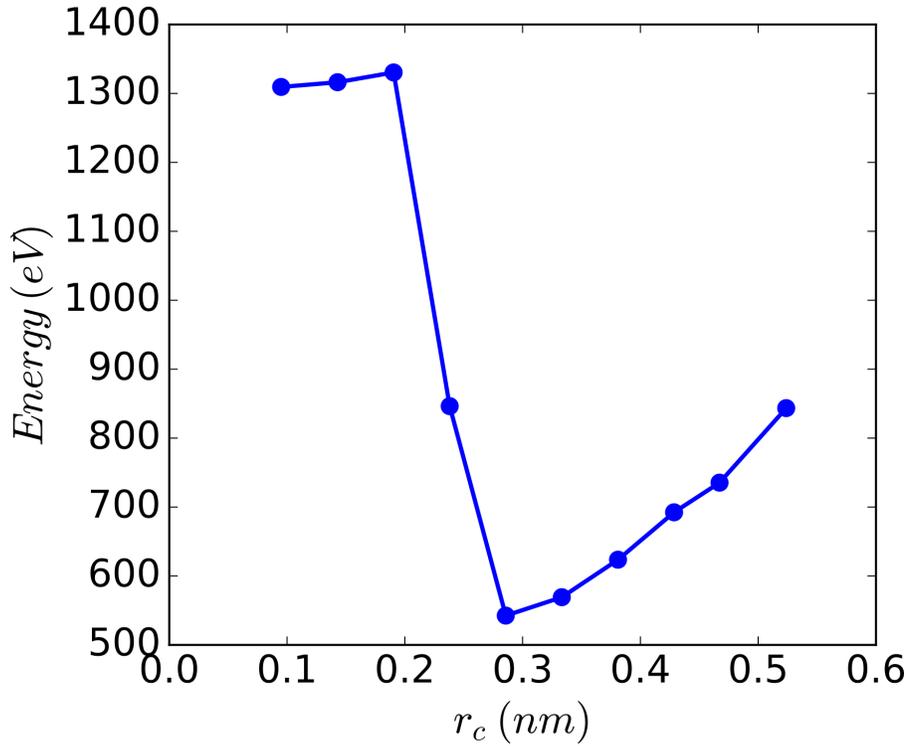}
	\caption{Excess of energy of the atomistic domains as a function of the  cut-off radius. The optimum cut-off radius to minimize the interface energy was 0.29 nm.}
	\label{Cutoff}
\end{figure}

The matrix and precipitate atoms were overlapped in the domain and a lozenge-shaped area inside the precipitate was selected and Mg atoms inside this area were removed. However, matrix and precipitate atoms were still overlapped at the interface. Thus, a cut-off radius (in the range 0.1 nm to 0.55 nm) was defined and all the atoms belonging to the Mg matrix within the cut-off radius of an atom of the precipitate were deleted. The atomistic models obtained with different cut-off radii were minimized using the conjugate gradient (CG) with periodic boundary conditions in all directions. The energy minimization was initially carried out at constant volume and subsequently at zero stress. The excess of energy of the relaxed domains in plotted in Fig. \ref{Cutoff} as a function of the cut-off radius. The minimum energy was obtained for a cut-off radius of 0.29 nm, which led to the semi-coherent interfaces with minimum energy. This structure was subjected to a thermal annealing to relieve the residual stresses by increasing the temperature up to 350 K in 100 ps and then remaining at this temperature during 200 ps within an NPT ensemble under periodic boundary conditions  and zero stress, followed by energy minimization using the CG. This final structure was then inserted into a larger Mg domain of $50 \times 25 \times 31$ nm$^{3}$, and then annealed within an NPT ensemble at 350 K and zero pressure for 10 ps. Finally, whole domain was relaxed using the CG algorithm.

In order to analyze the effect of the precipitate size on the dislocation/precipitate interaction mechanisms, two smaller $\beta$-Mg$_{17}$Al$_{12}$ with the same orientation and aspect ratio were created following the same procedure and inserted into the Mg matrix. Nevertheless, it was found that the lattice of the Mg$_{17}$Al$_{12}$ precipitate was destroyed after energy minimization, while the structure of the Mg matrix around the precipitate was not modified.
This behaviour was attributed to the small volume/surface ratio of these precipitates. More details can be found in Appendix \ref{Prec}.

\subsection{Atomistic simulation}\label{AtomSim}
All the atomistic simulations of the interaction of basal dislocations with $\beta$ precipitate were carried out using the open-source parallel molecular dynamics code LAMMPS \cite{lammps}. The modified embedded atom method (MEAM) interatomic potential for Mg-Al-Zn developed by Dickel \textit{et al.} \cite{DD18} was used in the simulations. This potential was selected among different interatomic potentials based on the accuracy to predict the properties of dislocations in Mg, the elastic constants of the Mg and $\beta$ precipitates as well as the interface energies between Mg and $\beta$ precipitates for the Burgers OR. The comparison of the properties and constants obtained with different Mg-Al interatomic potentials and first principles calculations can be found in Appendix \ref{DFT} and \ref{IP}.

A parallelepipedic domain of $50 \times 25 \times 31$ nm$^{3}$ along with the periodic array of dislocations and precipitates model \cite{OB03} was utilized. Periodic boundary conditions were applied along the $X$ and $Y$ directions   and non-periodic along the $Z$ direction.   The dimensions of the domain were chosen to minimize the image stresses due to the dislocation bowing during the simulations \cite{SC15}. An edge basal dislocation was introduced in the domain containing the precipitate by inserting a semi-plane of atoms into the model and applying the corresponding Volterra's displacement field \cite{atomsk}. The dislocation line was parallel to the $X$ axis and the Burgers vector was parallel to the $Y$ axis (Fig. \ref{Rot}).  The energy of the whole domain was minimized afterwards using CG and the perfect dislocation was split into two Shockley partials.

\begin{figure}[t]
	\centering
		\includegraphics[width=0.8\textwidth]{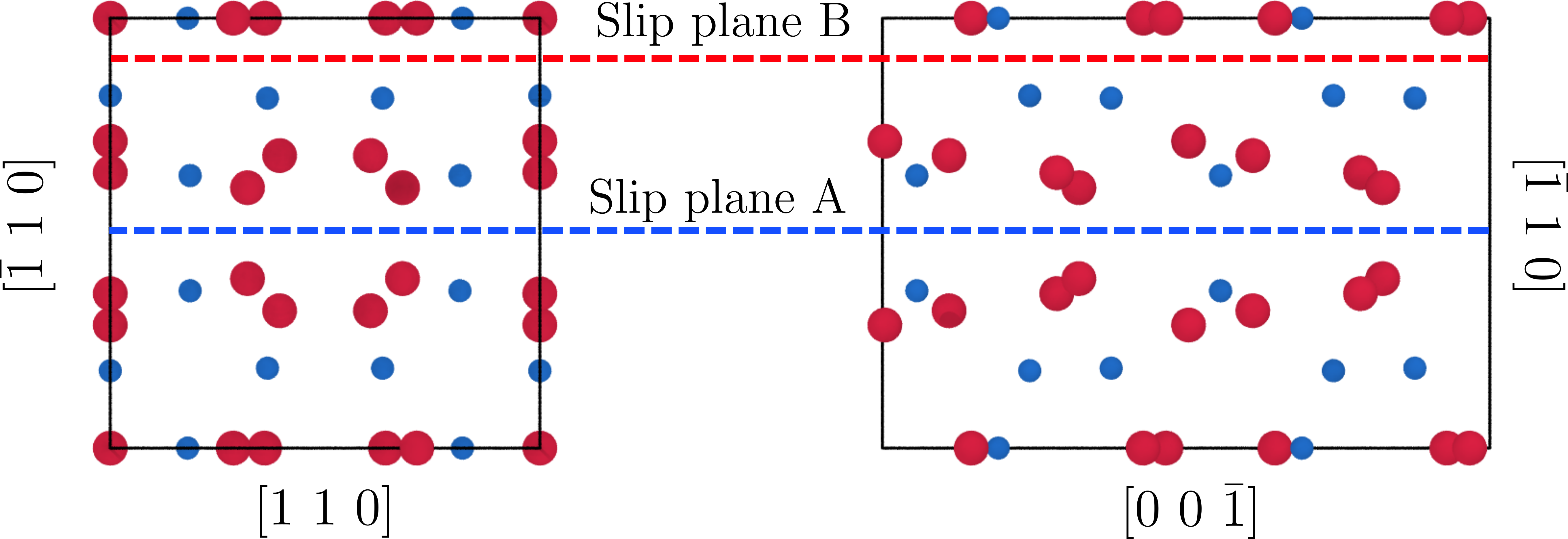}
	\caption{Plain views of the crystal structure perpendicular to the  (001) (left) and (110) (right) planes. The two slip planes (A and B) corresponding to the  [110](110) slip system (left) and [001](110) slip system (right) in the $\beta$ precipitate are indicated by broken red and blue lines.  These planes correspond to largest interplanar distances in the $[110]$ direction. Mg atoms are red and Al atoms blue.}
	\label{SlipPlanes}
\end{figure}

According to first-principles calculations the preferential slip plane for $\beta$ intermetallic is $<$110$>$, which provides the lowest energy barriers in the $\gamma$ energy curves \cite{xiao2013}. However, the $\gamma$ energy curves for the [110](110)$_\beta$ slip system also depend on the interplanar distances between pairs of equivalent planes due to the complex atomic structure of the precipitate. To account for this effect, the edge dislocation was placed in two different basal slip planes, A and B, of the Mg matrix that were separated by 0.8 nm (two atomic planes). These two slip planes are illustrated in the Fig. \ref{SlipPlanes}.

Two different types of atomistic simulations were carried out. The dislocation-precipitate interaction at 0 K was assessed by means of molecular statics (MS) simulations. Discrete shear displacements of 0.05 $\AA$ were successively applied in four atomic layers on the top surface of the domain, parallel to the slip plane, while four layers of atoms at the bottom surface remained fixed. The displacements in the top layers were constrained to one direction parallel to the slip direction. The energy was minimized after each displacement at constant volume. This procedure was repeated until the dislocation  overcame four times the precipitate. To evaluate the effect of the precipitate distance and size on the  CRSS  , further MS simulations were carried out. The distance between precipitates was evaluated by ranging the $Y$ axis to 20, 30, and 35 nm (apart from 25 nm that was the initial value). 

The interaction between the dislocation and the precipitate at finite temperature (10K and 300K) was evaluated by means of molecular dynamics (MD) simulations. The dynamic stabilization of temperature and stress was carried out using the NPT ensemble. A temperature ramp was applied to the system from 0K to the desired temperature in 10 ps. Afterwards, the temperature was kept constant during 15 ps using the same ensemble and the volume of the domain was allowed to expand by relaxing the normal stresses.  Then, the ensemble was change to NVT and shear displacement was prescribed to the atoms on the top surface while those at the bottom were fixed. The applied shear strain rate, $\dot{\gamma}$, was 1.3 $\times$ 10$^8$ s$^{-1}$, which was the lowest affordable with the computational resources available. Although it is much larger than the experimental ones, it should be noted that the mechanisms of dislocation/precipitate interaction (reported in the results section) were the same in the MS and MD simulations, indicating that were neither modified by strain rate nor temperature.   Stress, energy and atomic position data were computed and stored every 1 ps. The timestep in all MD simulations was 1 fs.

The visualization and analysis of the results were carried out by means of the open-source code OVITO \cite{OVITO}. Shear strains were calculated using atomic strain tensor algorithm \cite{Falk98,Shimizu07}. This method is based on the comparison of atomic displacement of the system with respect to a reference configuration. More specifically, the atomic deformation tensor is obtained from the relative atomic displacements of its neighbors (within a cutoff radius) and then the strain tensor was calculated from atomic deformation tensor. The dislocation extraction algorithm was used to evaluate dislocations inside the atomic domain \cite{Stukowski2010,Stukowski2012}. This algorithm creates a Delaunay tessellation of the atomic system and finds the edges that do not correspond to the perfect lattice. Then, the dislocations are identified by the Burgers circuit around the corresponding atoms of the non-perfect edges.

\section{Results and discussion}
\subsection{Dislocation/precipitate interaction mechanism at 0K}
\begin{figure}
	\centering
		\includegraphics[width=\textwidth]{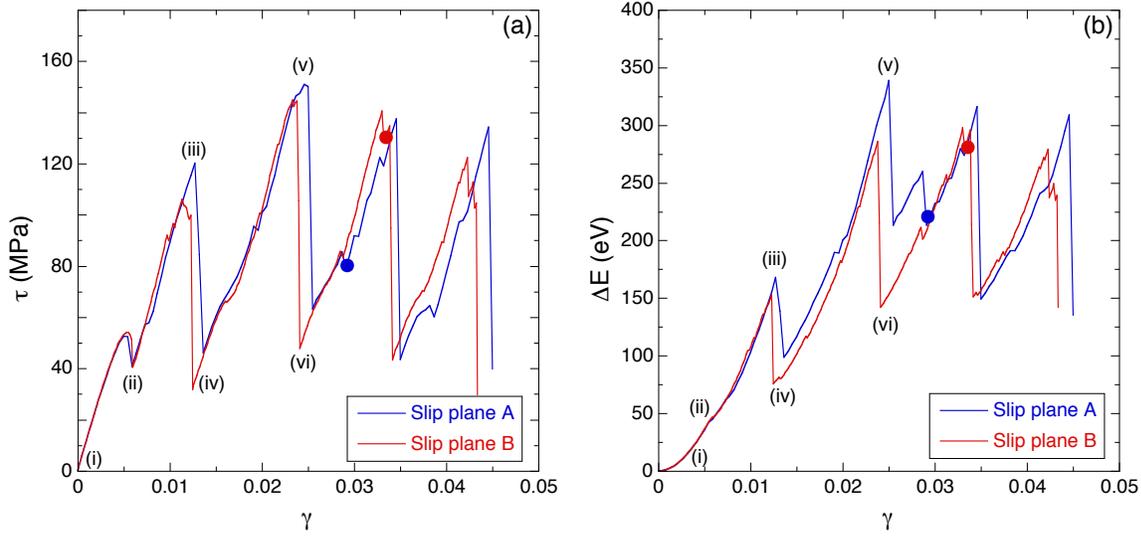}
	\caption{Molecular statics simulations of dislocation/precipitate interaction. (a) Shear stress {\it vs.} strain curves ($\tau - \gamma$) corresponding to slip planes A and B. (b) Stored energy {\it vs.} strain curves ($\Delta E - \gamma$)  corresponding to slip planes A and B. (c) Dislocation/precipitate interactions along slip plane A. (d) Dislocation/precipitate interactions along slip plane B. The blue and red solid circles in (a) and (b)   denote the exact moment at which the precipitate is sheared by the first dislocation along the slip planes A and B, respectively.  }
	\label{MS}
\end{figure}

The shear stress - shear strain ($\tau$- $\gamma$) and the stored energy - shear strain curves ($\Delta E - \gamma$) obtain by MS simulations are depicted in Figs. \ref{MS}(a) and (b), respectively. Similar behaviors, either in stress and store energy \textit{vs} strain, were found when the basal dislocations slipped along the A and B planes and they will be discussed together. The stress and the energy stored increased initially with $\gamma$ due to the elastic interaction of the dislocation with the precipitate   and a local minimum in the stress was reached at $(ii)$ due to the attraction of the leading partial to the interface. The dislocation was pushed towards the precipitate as the applied strain increased and the CRSS at which the dislocation overcame the precipitate was reached at $(iii)$. It was followed by strong reduction in the shear stress and in the energy stored, marked as ($iv$), as the first dislocation overcame the precipitate.

Afterwards, the second dislocation appeared due to periodic boundary conditions and the shear stress and the energy stored increased as the new dislocation approached the precipitate (from $(iv)$ to $(v)$ in Fig. \ref{MS}(a) and (b), respectively). By-passing the precipitate led to another marked reduction in the stress and stored energy, marked as ($vi$), and this process was repeated every time a new dislocation overcame the precipitate. It is worth noting that  the CRSS to overcome the precipitate did not increase (except between the first and the second dislocation) with the number of dislocations overcoming the precipitate, as it would be expected for an Orowan-type mechanism. This behavior suggests that the plastic strain was somehow transferred to the precipitate.

\begin{figure}
	\centering
		\includegraphics[width=0.9\textwidth]{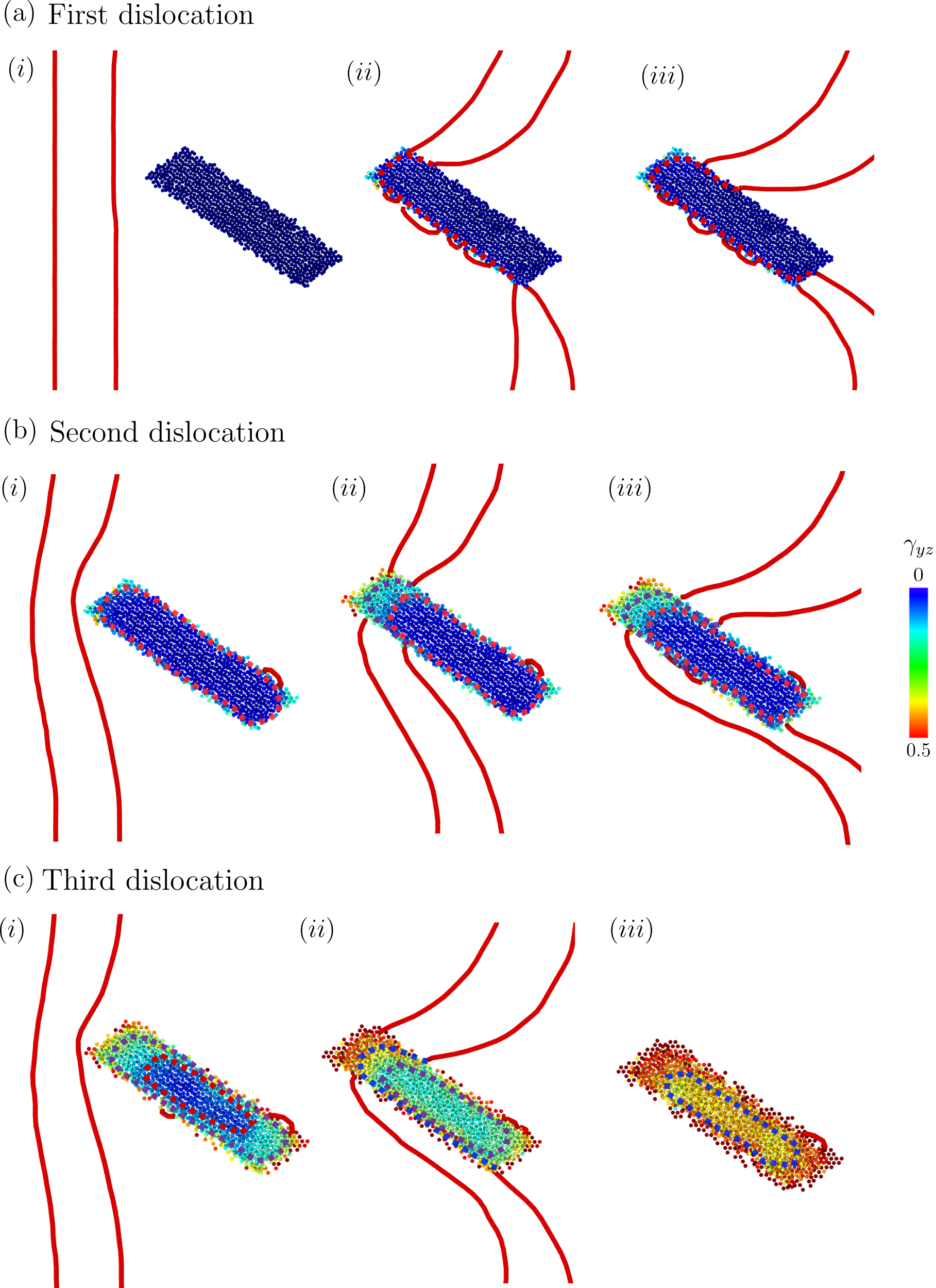}
	\caption{Details of the interaction of the partial dislocations (red lines) with the Mg$_{17}$Al$_{12}$ precipitate, which is sheared along the (110) crystallographic slip plane. (a) First dislocation. (b) Second dislocation. (c) Third dislocation. The contour plot of the precipitate section shows the shear strain, $\gamma_{yz}$. Only the atoms of the (110) crystallographic slip plane of the precipitate, which is the continuation of the slip plane A, are shown.}
	\label{Partial}
\end{figure}

The details of the interaction of the successive dislocations with the precipitate for the slip system A are depicted in Fig.  \ref{Partial}, where the partial dislocations are depicted as red lines and only the atoms of the (110) crystallographic slip plane A in the precipitate are shown. The colour of the atoms stands for the shear strain in the slip plane, $\gamma_{yz}$. The first dislocation was initially attracted by the ($\bar 2 \bar 1$10)$_{Mg}$/($2\bar 2 \bar 1$)$_{\beta}$ short interface, and was able to penetrate slightly into the precipitate, Fig. \ref{Partial}(a)$(ii)$. However, precipitate shearing was not possible and the partial dislocations propagated along the interface (Fig. \ref{Partial}(a)$(iii)$). Thus, the dislocation overcame the precipitate by the formation of an Orowan loop, but a clear loop in the Mg matrix around the precipitate was not found. The loop seemed to be formed into the precipitate, as suggested by the the larger strain at the edges of the precipitate in Fig. \ref{Partial}(b)$(i)$, but it was not easy to identify due to the complexity of the precipitate lattice.

The interaction of the second dislocation with the precipitate is depicted in Fig. \ref{Partial}(b)$(i)$ to $(iii)$. Initially, there was a repulsive interaction between the partial dislocations and the Orowan loop within the precipitate, as shown in $(i)$. Further deformation led to the penetration of the leading partial $(ii)$ and, afterwards, of the trailing partial $(iii)$ into the precipitate.  Moreover, a gradient in the atomic shear strain along the slip plane appeared in the precipitate, suggesting that the Orowan loop formed by the first dislocation has propagated further inside the precipitate.  Finally, the second dislocation overcame the precipitate leaving two Orowan loops, which have penetrated deeper into the precipitate, as depicted in Fig. \ref{Partial}(c)$(i)$.   

Finally, the interaction of the third dislocation with the precipitate is depicted in Fig. \ref{Partial}(c). It followed the pattern of the first and second dislocations with the leading partial approaching the precipitate through the short interface, Fig. \ref{Partial}(c)$(i)$. As soon as it entered the precipitate, the first Orowan loop within the precipitate collapsed and sheared the precipitate, Fig. \ref{Partial}(c)$(ii)$. This instant corresponds to the blue dot in Figs. \ref{MS}(a) and (b). The propagation of the leading and trailing partials along the interface finally led to the collapse of the Orowan loop induced by the second dislocation, Fig. \ref{Partial}(c)$(iii)$, and the precipitate was sheared again.

Successive dislocation passes led to close values of shear stress and stored energy, as Figs. \ref{MS}(a) and (b) depicts, reaching a maximum value with the second dislocation and the dislocation overcoming mechanism did not change in the subsequent passes, although the disorder at the interface increased due to the absorption the previous dislocations. The slight differences in the $\tau-\gamma$ and $\Delta E - \gamma$ curves between slip planes A and B can be attributed to the differences in the interface structure and also to the mechanisms of deformation in the precipitate. It is interesting to notice that this mechanism of dislocations gliding along the matrix/precipitate interface has been recently reported during {\it in situ} nanoindentation experiments in the transmission electron microscope in Mg-Nd alloys \cite{HAM18}. 

\begin{figure}[t]
	\centering
	\includegraphics[width=0.8\textwidth]{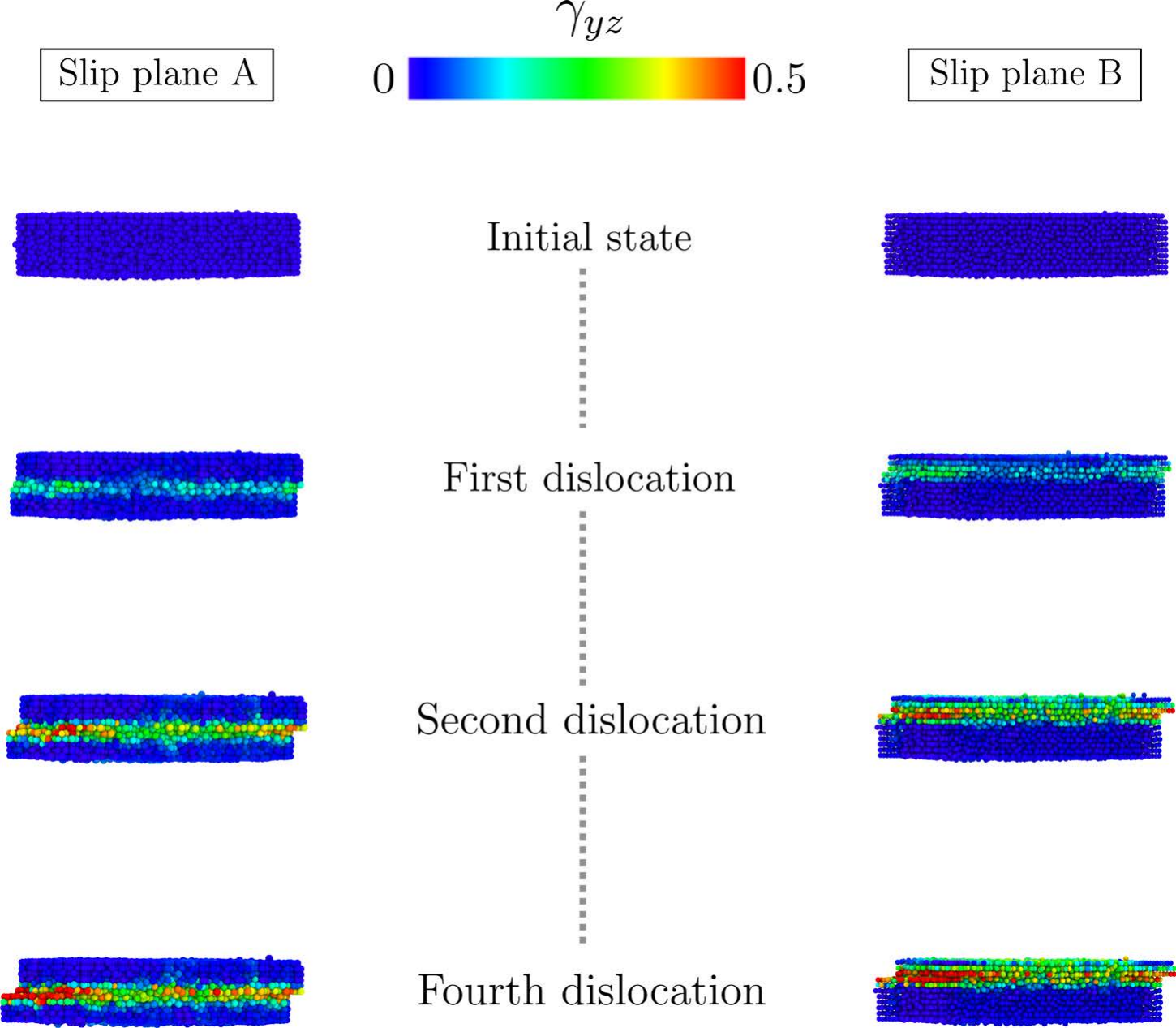}
	\caption{Contour plots of the shear strain parallel to the slip plane ($\gamma_{yz}$) in the cross-section of the precipitate as a function of the number of dislocations that bypass the precipitate. The results on the left correspond to  slip plane A and those on the right to slip plane B. }
	\label{SP}
\end{figure}

To better ascertain the overcoming mechanism, the contour plot of the shear strain in the cross-section of the precipitate is shown in Figs. \ref{SP} for slip planes A and B. The precipitate was strain-free at the initial state but shear strains were very localized around the slip plane (particularly for the precipitate sheared along the slip plane A) after the first dislocation has overcome the precipitate. The boundary region affected by the large shear strain, increased with the number of dislocations bypassing the precipitate (Fig. \ref{Partial})  but the thickness of this shear-deformed zone remained very thin (just a few atomic planes) and, finally, the precipitate was sheared. This process is compatible with the progressive shearing of the precipitate by the successive dislocation loops. The first dislocation penetrated the precipitate but could not propagate further and remained close to the surface. The following dislocation loop pushed the initial dislocation loop further into the precipitate and this process was repeated until the first dislocation loop completely sheared the precipitate and was annihilated. This point is associated with a reduction of the shear stress and energy storage, as shown by the blue and red solid circles in Figs. \ref{MS}(a) and (b). Further deformation led to progressive shearing of the precipitate by the successive dislocations. The whole process can be observed in the movie PrecipitateShearing.mov in the supplementary material.

 These   static simulations depict a dislocation/precipitate interaction mechanism that it is a mixture of Orowan looping and precipitate shearing. The initial dislocations overcome the precipitate by the formation of an Orowan loop that penetrates the precipitate but it was not able to shear it because of the differences in the structure and the high  CRSS   necessary to move dislocations in the precipitate.  Another Orowan loop was introduced into the precipitate after the next dislocation bypassed the precipitate, and the first Orowan loop is pushed further into the precipitate. This process continued until the first dislocation loop completely sheared the precipitate and was annihilated. This mechanism was repeated as more dislocations overcome the precipitate, and the precipitate was eventually sheared. The detailed process of precipitate shearing is shown in Appendix \ref{PS}

\subsection{Dislocation/precipitate interaction mechanisms at finite temperatures}
\begin{figure}
	\centering
		\includegraphics[scale=0.8]{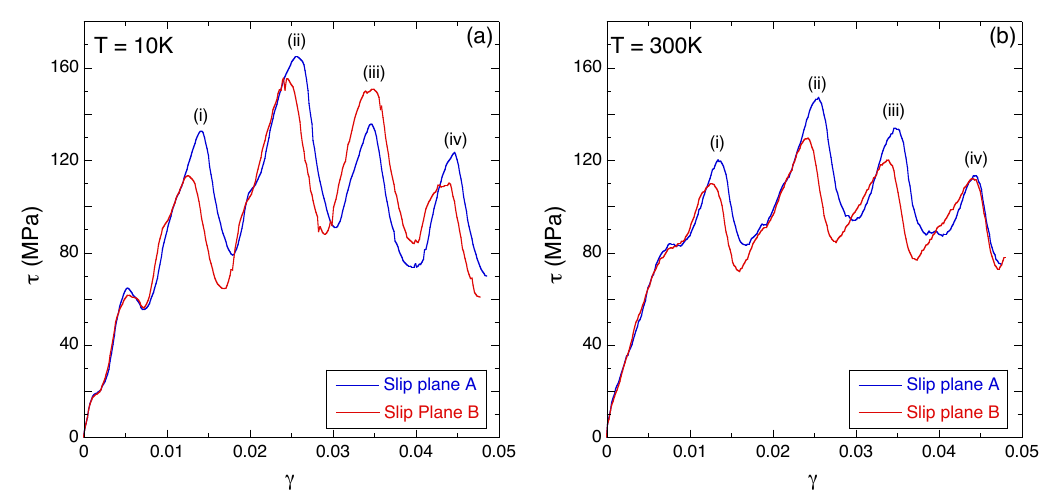}
	\caption{Shear stress {\it vs.}strain curves ($\tau - \gamma$) corresponding to slip planes A and B obtained by molecular dynamics. (a) T = 10 K. (b) T = 300 K. The peaks marked (i), (ii), (iii), and (iv) indicate when the successive dislocations overcome the precipitate.}
	\label{MD}
\end{figure}

The effect of temperature on the dislocation/precipitate interactions was analyzed by means of MD calculations. Simulations were carried out for dislocations located in the planes A and B and the corresponding shear stress {\it vs.} shear strain curves  are depicted in Fig. \ref{MD}(a) and (b) at 10 K and 300 K, respectively. The mechanical response at both temperatures was very similar and these curves were also very close to those obtained at 0 K (Fig. \ref{MS}), indicating that the energy barrier to bypass the precipitates was very large and it was not influenced by the temperature in the MD simulations. The curves present successive peaks in the shear stress (marked (i), (ii), (iii) and (iv)), which correspond to the bypass of the precipitate by the successive dislocations. The maximum value of the shear stress was attained after the second dislocation overcome the precipitate and its magnitude was similar to the one reported at 0 K in Fig. \ref{MS}. Afterwards, the precipitate was sheared following a mechanism equivalent to the one reported above (Figs. \ref{slipA} and \ref{slipB}) and the  CRSS   necessary to overcome the precipitate by the following dislocations decreased.

The mechanisms of dislocation/precipitate interaction reported above are different from those observed in other precipitate-strengthened alloys. Precipitate shearing by dislocations is normally found in the case of coherent precipitates ;   the matrix dislocation enters the precipitate through the coherent interface and slips along the most suitable slip plane. This phenomenon is most clear when the matrix and the precipitate share the same crystallographic lattice (for instance, $\gamma'$ precipitates in a $\gamma$ matrix in Ni-based superalloys \cite{VDR09} or Guinier-Preston zones in Al-Cu alloys \cite{EMS19}) but it has also been reported in the case of basal dislocations with $\beta'$ precipitates in Mg-RE alloys \cite{SM18, BWS18}. Under these circumstances, the  CRSS   necessary to shear the precipitate depends on the coherency strain, the modulus mismatch, the chemical strengthening and, in the case of ordered precipitates, on the energy penalty due to the formation of antiphase boundaries \cite{N97}. On the contrary, dislocations cannot enter the precipitate in the case of incoherent interfaces, and the obstacle to the dislocation motion is overcome by the formation of an Orowan loop around the precipitate. The shear stress necessary to overcome the precipitate can be calculated using dislocation line tension models \cite{O48, BKS73} and the interaction of successive dislocations with the Orowan loops around the precipitates leads to a strong hardening. 

Nevertheless, shearing of non-coherent precipitates by basal dislocations in Mg-Zn \cite{WS15, AL19},  Mg-Al alloys \cite{VKM17, CCP19} and Mg-Nd alloys \cite{ZWZ18, HYQ19} has been recently reported in several experimental studies. The common feature of all these investigations is that the Mg basal plane is parallel to one crystallographic plane of the precipitate: $(0001)_{Mg} \| (11\bar2 0)_{\betaÕ}$ in Mg-Zn, $(0001)_{Mg} \| (110)_{\beta}$ in Mg-Al, and $(0001)_{Mg} \| (110)_{\beta_1}$ in Mg-Nd. This crystallographic correspondence between the basal plane of Mg and closely packed crystallographic planes of the precipitates occurs spontaneously during precipitate nucleation. Even though the favourable matrix/precipitate interfaces may be incoherent (due to the different crystallographic structure and the mismatch in the lattice parameters), basal dislocations in the Mg matrix can glide into the precipitate without changing the slip plane. Of course, the stress necessary to shear the precipitate is much higher than that for basal slip in Mg and several basal dislocations have to pile-up at the interface before the precipitate is sheared. The presence of dislocation pile-ups before precipitate shearing has also been experimentally reported in Mg-Al \cite{CCP19} and Mg-Nd \cite{HYQ19}, supporting our atomistic results. It should be also noted that precipitate shearing in Mg alloys by either prismatic or pyramidal dislocations was not observed because of the lack of crystallographic continuity between the corresponding slip planes in the matrix and in the precipitate \cite{BWS18}. Thus, the strength of Mg alloys is limited by the low  CRSS   for basal slip, which is always the main plastic deformation mechanism. Precipitate shearing by basal dislocations limits the strengthening provided by precipitates (hardening due to the interaction of dislocations with Orowan loops around the precipitates is not active) and, thus, hinders precipitation hardening in Mg alloys.

Finally, it is worth noting the importance of creating a low energy matrix/precipitate interface with the proper interface orientations to analyze the dislocation/precipitate interactions. If these details are not included in the atomistic model, the interaction of the dislocations with the interface as well as the residual stresses created to introduce the precipitate into the matrix may lead to the  appearance   of spurious deformation mechanism in the atomistic simulations.
 
\subsection{Comparison with theoretical models}

\begin{figure}[t]
	\centering
		\includegraphics[scale=1.0]{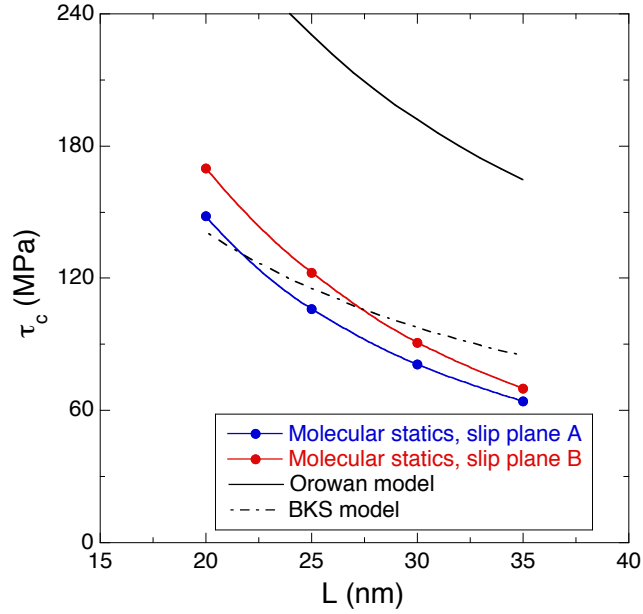}
\caption{Resolved shear stress ($\tau_c$) to form an Orowan loop as a function of the distance between precipitates, $L$. Results obtained with molecular statics simulations, Orowan model ($\tau_c = \mu b /L$) and BKS model (eq. (\ref{eq:BKS})) are plotted for comparison.}
	\label{Orowan}
\end{figure}

The results presented above showed that the dislocations initially overcome the precipitates by the formation of an Orowan loop and it is interesting to compare the atomistic predictions of the resolved shear stress, $\tau_c$ after the first dislocation has overcome the precipitate with those of the Orowan model \cite{O48}. To this end, molecular statics simulations were carried out using simulations boxes with different widths along the $X$ axis (from 20 nm to 35 nm) while the precipitate dimensions and orientation were not modified. $\tau_c$ in the simulations was equal to the stress necessary to overcome the precipitate by the first dislocation, which was always below the  CRSS . These results are compared in Fig. \ref{Orowan} with the predictions of the Orowan model, $\tau_c = \mu b/L$, where $\mu$ is the shear modulus of the Mg crystal parallel to the basal plane (18 GPa), $b$ the Burgers vector of basal dislocations (0.32 nm) and $L$ the distance between precipitates along the dislocation line, which is equal to the simulation box width.

It was found that the Orowan model overestimated $\tau_c$ along both slip planes A and B (Fig. \ref{Orowan}) and the differences may be attributed to the significant bowing of the dislocations to overcome the precipitate (Figs. \ref{MS}c and d). The energy associated with the interaction of opposite dislocation segments can provide an important contribution to  $\tau_c$ that can be rationalized by means of the Bacon-Kocks-Scattergood (BKS) model \cite{BKS73}. In this model, the resolved shear stress necessary to overcome the precipitate takes into account the interaction between dislocation segments and scales with the natural logarithm of the harmonic mean between the precipitate diameter $D$ and the distance between precipitates $L$ according to

\begin{equation}
\tau_c = \frac{\mu b}{2\pi A L} \bigg(\ln\bigg[ \frac{b}{D} + \frac{b}{L}   \bigg]^{-1} + B\bigg)
\label{eq:BKS}
\end{equation}

\noindent where $A$ = 1 for edge dislocations and $B$ an adjustable parameter. In this case, $D$  = 10.6 nm was taken as the projection of the precipitate on the dislocation line. The predictions of the BKS  (with $B$ = 0.075) are plotted in Fig. \ref{Orowan} together with results of the MS simulations. They are in good agreement and indicate that the BKS model can provide reasonable predictions of the stress necessary to form an Orowan loop around the $\beta$ precipitates in Mg-Al alloys.

The  CRSS   needed to overcome the precipitate was controlled in the MS and MD simulations by the stress necessary to shear the precipitate. This stress was attained by the pile-up of two dislocations around the precipitate. Further deformation did not lead, however, to the accumulation of dislocations at the precipitate and to  strain hardening, because the precipitate was sheared. Thus, this dislocation/precipitate interaction mechanisms limits the strain hardening capability of Mg alloys as compared to metallic alloys containing precipitates that cannot be sheared by dislocations. Moreover, it should be noted that precipitate shearing has been experimentally reported in $\beta$-Mg$_{17}$Al$_{12}$ precipitates much larger than those analyzed in the atomistic simulations in this investigation \cite{VKM17, CCP19}. 

\section{Conclusions}
The interaction between edge basal dislocations and $\beta$-Mg$_{17}$Al$_{12}$ precipitates was studied using atomistic simulations. A strategy was developed to insert a lozenge-shaped Mg$_{17}$Al$_{12}$ precipitate with Burgers orientation relationship within the Mg  matrix in an atomistic model. To this end, the precipitate with the right shape and orientation relationship (according to the experimental data) was introduced into the Mg matrix. Afterwards,  all the atoms belonging to the Mg matrix within the cut-off radius of an atom of the precipitate were deleted and the optimum cut-off radius was obtained by energy minimization. Finally, the simulation box was annealed to relieve the remaining residual stress. In this way, the incoherent matrix/precipitate interfaces in the atomistic model were close to stable, minimum energy configurations.
 
Molecular statics and dynamics simulations of the dislocation precipitate interaction showed similar mechanisms. The dislocation bypassed the precipitate by the formation of an Orowan loop, that entered the precipitate but it was not able to progress further until more dislocations bypass the precipitate and push the initial loop to shear the precipitate along the (110) plane, parallel to the basal plane of Mg. This process was eventually repeated as more dislocations overcome the precipitate. These mechanisms were in agreement with the experimental evidences \cite{VKM17, CCP19} and indicate  that precipitate shearing by basal dislocations in Mg-Al alloys is favored because the  (0001) Mg basal  plane is parallel to the (110) crystallographic plane of the precipitate. 

The resolved shear stress to form an Orowan loop around the precipitate was in agreement with the predictions of the Bacon-Kocks-Scattergood model that takes into account the interaction between dislocation segments. The mechanisms of dislocation/precipitate interaction as well as the resolved shear stress to form an Orowan loop were independent of the temperature in the range of temperatures and strain rates explored by the molecular dynamics simulations. Finally, the details of the shearing mechanism of the precipitate depended on the particular (110) plane in which the shear strain was localized but they did not significantly influence the  CRSS .

\section{Acknowledgments}

This investigation was supported by the European Research Council under the European UnionÕs Horizon 2020 research and innovation programme (Advanced Grant VIRMETAL, grant agreement No. 669141). The computer resources and the technical assistance provided by the Centro de Supercomputaci\'on y Visualizaci\'on de Madrid (CeSViMa) are gratefully acknowledged. Additionally, the authors thankfully acknowledge the computer resources at Finisterrae and the technical support provided by CeSGa and Barcelona Supercomputing Center (project QCM-2017-3-0007). Finally, use of the computational resources of the Center for Nanoscale Materials, an Office of Science user facility, supported by the U.S. Department of Energy, Office of Science, Office of Basic Energy Sciences, under Contract No. DE-AC02-06CH11357, is gratefully acknowledged. EM used resources provided by the Los Alamos National Laboratory (LANL) Institutional Computing Program. LANL is operated by Los Alamos National Security, LLC, for the National Nuclear Security Administration of the U.S. DOE under contract DE-AC52-06NA25396.

\section{Appendix}

\setcounter{table}{0}
\setcounter{figure}{0}
\setcounter{section}{0}
\renewcommand{\thetable}{A\arabic{table}}
\renewcommand{\thefigure}{A\arabic{figure}}
\renewcommand{\thesection}{A.\arabic{section}}

\section{ Stability of small $\beta$-Mg$_17$Al$_12$ precipitates in the Mg matrix }\label{Prec}

\begin{figure}
	\centering
	\includegraphics[width=\textwidth]{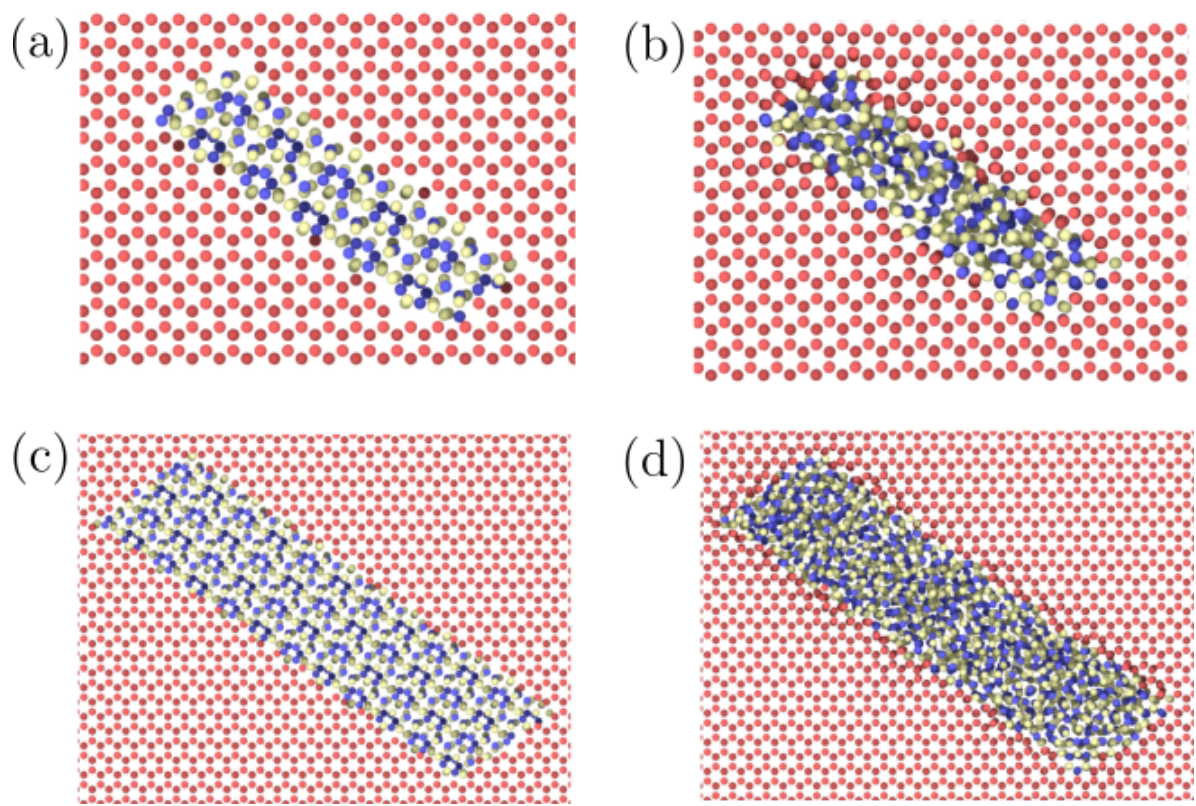}
	\caption{(a) Initial atomic structure of the 4 $\times$ 1 $\times$ 2.8 nm$^3$ $\beta$-Mg$_{17}$Al$_{12}$ precipitate embedded the Mg matrix. (b) {\it Idem} after energy minimization to create interfaces with minimum energy. (c)  Initial atomic structure of the 8 $\times$ 2 $\times$ 2.8 nm$^3$ $\beta$-Mg$_{17}$Al$_{12}$ precipitate embedded the Mg matrix. (d) {\it Idem} after energy minimization to create interfaces with minimum energy.}
	\label{Beta_Small}
\end{figure}

  The stability of  $\beta$-Mg$_{17}$Al$_{12}$ precipitates in the Mg matrix was analyzed as a function of its size. To this end, two smaller models of precipitate with the same aspect ratio (4:1) parallel to the basal plane and Burgers OR  were created. The dimensions of these precipitates were 4 $\times$ 1 $\times$ 2.8 nm$^3$ and 8 $\times$ 2 $\times$ 2.8 nm$^3$. They were inserted in the Mg matrix and the optimum cut-off radius, which led to the semi-coherent interfaces with minimum energy, was obtained as indicated in section 3.1.  Nevertheless, the original precipitate structure (that is shown in Figs. \ref{Beta_Small}(a) and (c) for both precipitate sizes) was destroyed during the process to minimize the energy of the interfaces, as shown in Figs. \ref{Beta_Small}(b) and (d). This dramatic change of the precipitate structure can be attributed to the large surface/volume ratio of these small precipitates where the interface energies determine the actual atomic arrangement within the precipitate.  

\section{Density Functional theory calculations}\label{DFT}

The first principles calculations in the present study were carried out using  the Quantum Espresso plane-wave pseudopotential code \cite{QE}. The Perdew-Burke-Erzenhof (PBE) approach \cite{GGA} was used to evaluate the exchange-correlation energy, within the Generalised Gradient Approximation (GGA). Ultrasoft pseudopotentials were used to reduce the basis set of plane wavefunctions used to describe the real electronic functions \cite{USP}.  After careful convergence tests, a cutoff of 36 Ry (490 eV) was found to be sufficient to reduce the error in the total energy to less than 1 meV/atom. A separation of 0.03 $\AA^{-1}$ in the k-point grid  was employed for the integration over the Brillouin zone according to the Monkhorst–Pack scheme \cite{MP}. The elastic constants were determined by applying a given strain and calculating the stress, as the unit cell was kept fixed and only the internal coordinates were allowed to relax. Deformation of each unit cell was carried out in different orientations taking into account the symmetries of each crystal structure. Six strain steps (varying from -0.003 to 0.003) were used for each orientation to obtain a reliable linear fit of the stress-strain relationship. 

\section{Selection of the interatomic potential for MS/MD}\label{IP}

There is a number of interatomic potentials for the Mg-Al system, namely: Liu's  Embedded Atom  Method  (EAM) (alloy type) \cite{liu1996eam}, Mendelev's EAM Finnis-Sinclair type  \cite{MAR09},  Jelinek's  MEAM potential \cite{jelinek}, and Kim's \cite{kim} and Dickel's \cite{DD18} 2NN MEAM potentials. Out of these five potentials, Jelinek's potential predicts a positive formation energy of 49 meV/atom for the Mg$_{17}$Al$_{12}$ phase, as compared to -30 meV/atom obtained from density functional theory (DFT) calculations. Preliminary molecular statics simulations were carried out to analyze the dislocation/precipitate interaction in Mg-Al alloys with the aforementioned potentials. In the case of Liu's potential, a spurious jog developed when the dislocation approached the precipitate. The dislocation inside the precipitate climbed to the precipitate/matrix interface and slipped along the interface without shearing the precipitate. This behaviour has been reported already in \cite{LLH14}. 

\begin{table}
 \begin{center}
    \caption{Peierls stress ($\sigma_p$) of an $<a>$ edge basal dislocation (MPa), stacking fault energy (SFE) of 1/3 $<1010>$ (mJ/m$^2$) and elastic constants of Mg (GPa). } 
    \label{tab:lp_ecs}
    \begin{tabular}{l|c|c|c|c|c|c|c|c|c|c|c} \hline
      Method & $\sigma_p$& SFE & $C_{11}$ & $C_{12}$ & $C_{33}$ & $C_{13}$ &  $C_{44}$ & $C_{66}$  \\
      \hline
		Mendelev \cite{MAR09} & 0.33 & 44.2 & 61.9 & 26.2 & 68.2 & 22.1 & 18.2 & 18  \\
		Liu \cite{liu1996eam} & 24.5 & 54.2 & 68.2 & 26.1 & 71.3 & 16.1 & 12.8 & 21.3  \\
		Kim  \cite{kim} & 3.03 & 29.6 &  62.9 & 26.32 & 69.9 & 21.2 & 17.2 & 18.2 \\
		Dickel  \cite{DD18} & 0.25 & 23 &  64.4 & 25.3 & 70.9 & 20.3 & 18 & 19.5 \\
		DFT  & & 36 &  59.8 & 22.6 & 62.5 & 25.6 & 16.8 & 16.7  \\
      \hline
    \end{tabular}
  \end{center}
\end{table}

\begin{table}
 \begin{center}
    \caption{Elastic constants of  $\beta$-Mg$_{17}$Al$_{12}$ (GPa).} 
    \label{tab:lp_ecsb}
    \begin{tabular}{l|c|c|c|c|c|c|c|c|c|c|c}  \hline
      Method &  $C_{11}$  & $C_{12}$ &  $C_{44}$ \\
      \hline
		Mendelev \cite{MAR09}  & 137.8 & 46.5 & 28 \\
		Liu \cite{liu1996eam}  & 135.3 & 47.8 & 23.2 \\
		Kim  \cite{kim}  & 84 & 32.5 & 13.7\\
		Dickel  \cite{DD18} & 89.4 & 33.5 & 19.6 \\
		DFT  & 83.2 & 29.6 & 17.1  \\
      \hline
    \end{tabular}
  \end{center}
\end{table}

The elastic constants of Mg and $\beta$-Mg$_{17}$Al$_{12}$  precipitates obtained by DFT as well as with the different interatomic potentials using MS are shown in Tables \ref{tab:lp_ecs} and \ref{tab:lp_ecsb}. It can be easily seen that only the results obtained with Kim's and Dickel's potentials were close to the DFT  results and to the experimental values for Mg \cite{HLD14}. The next step of the evaluation of the candidate potentials was to determine the Peierls Stress of an $<a>$ edge basal dislocation and the stacking fault energy in the basal plane of Mg (Table \ref{tab:lp_ecs}). Regarding the Peierls stress, only the result from Dickel's potential (0.25 MPa) is in quantitative agreement with experimental results reported in the literature \cite{YASI_PS}, while both Kim's and Dickel's potential underestimate slightly the staking fault energy.

\begin{table}
 \begin{center}
    \caption{Interface energies (mJ/m$^2$) of the different orientations between Mg and $\beta$-Mg$_{17}$Al$_{12}$ in the Burgers OR.} 
    \label{tab:int_en}

    \begin{tabular}{c|c|c|c} 
    
     & $(0001)_{Mg}\|(110)_{\beta}$ &  $(\overline 1 \overline 1{2}0)_{Mg}\|(001)_{\beta}$ \\
      \hline
		Dickel  \cite{DD18} & 166 &  459 \\
		Kim  \cite{kim} & 165 &  459 \\		
		DFT  &  246  & 348 \\
      \hline
    \end{tabular}%
  \end{center}
\end{table}

\begin{figure}
\centering
\includegraphics[scale=0.6]{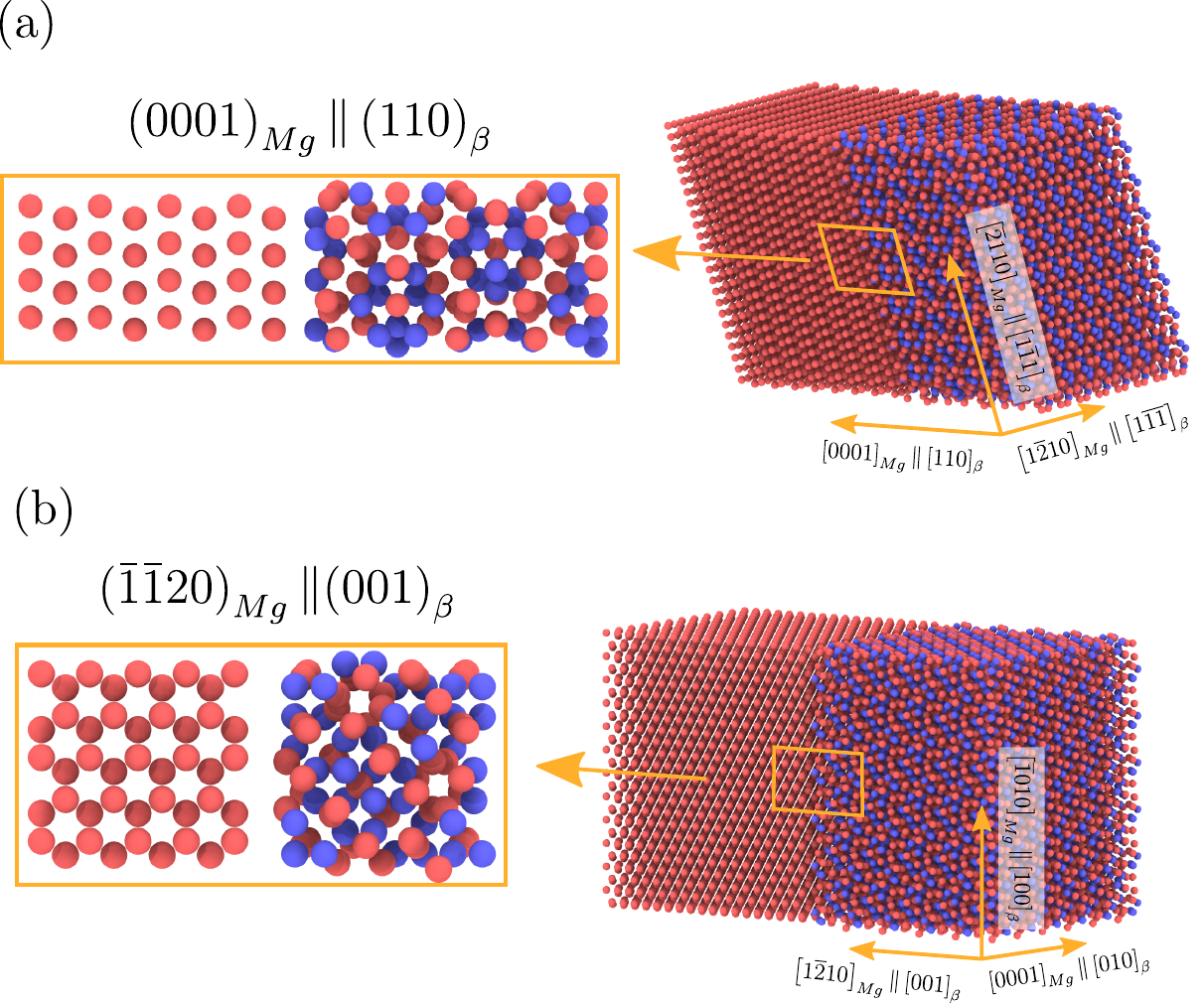}
\caption{Structures of the interfaces between Mg and $\beta$-Mg$_{17}$Al$_{12}$ precipitate in the Burgers OR. (a) $(0001)_{Mg}\|(110)_{\beta}$ interface. (b) $(11\overline{2}0)_{Mg}\|(001)_{\beta}$ interface.}
\label{fig:int}
\end{figure}

Another important property that needs to be addressed when investigating dislocation-precipitate interaction is the interface energy. The two main interfaces between the Mg matrix and the $\beta$ precipitate in the Burgers OR are $(0001)_{Mg}\|(110)_{\beta}$ and $(\overline 1 \overline 1{2}0)_{Mg}\|(001)_{\beta}$ and they are shown in Fig \ref{fig:int}. The energies of these interfaces were calculated using DFT and MEAM potentials from supercell calculations and the strategy to determine the interface energies is detailed in \cite{RV18}.  No different terminating surfaces were evaluated, as the purpose of these calculations was to benchmark and evaluate the precision of the interatomic potentials against DFT. The results are shown in Table \ref{tab:int_en} and both Dickel and Kim MEAM potentials provided equivalent results, which were cin qualitative agreement with the DFT results.

Taking into account all the previous results, Dickel's potential was selected to study the dislocation/precipitate interaction. One additional advantage of this potential is that it can be used to model systems with Zn content (up to 2 at. \%) \cite{DD18}, opening the way to analyze the behavior of ternary alloys. 

\section{  Shearing of the $\beta$ precipitate by dislocations along the slip planes A and B.}\label {PS}
\begin{figure}[t]
	\centering
	\includegraphics[width=0.9\textwidth]{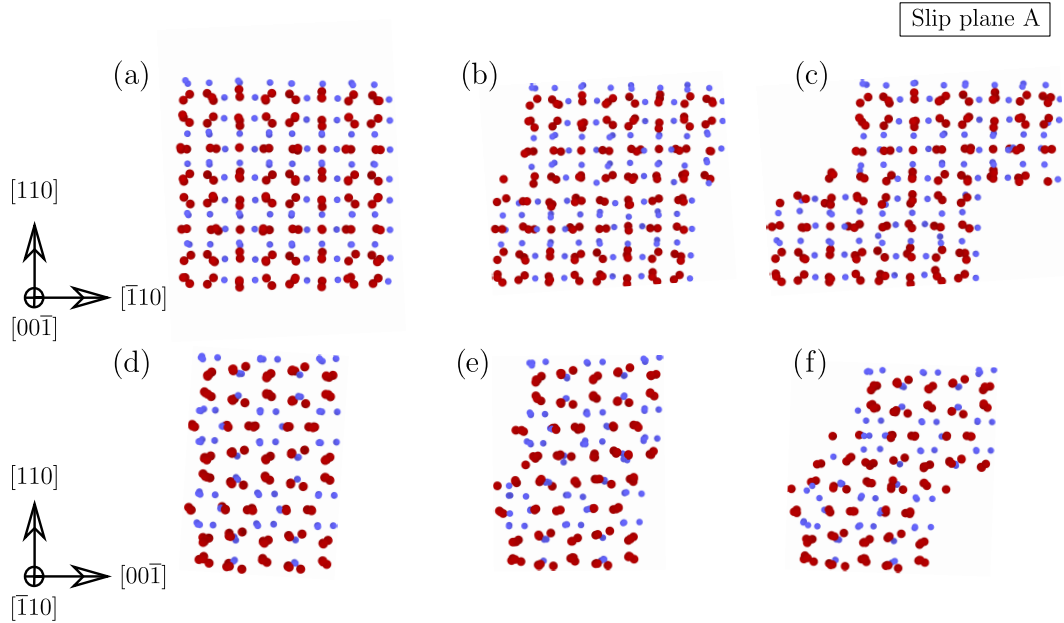}
	\caption{Shearing of the precipitate along the slip plane A. (a) and (d) Initial cross-sections.  (b) and (e)  cross-sections for the strain marked by a  blue circle in Figs. \ref{MS}(a) and (b)). (c) and (f) Cross sections  at the end the molecular statics simulations after four dislocations have bypassed the precipitate. The crystallographic directions in this figure correspond to the $\beta$ precipitate.}
	\label{slipA}
\end{figure}

The cross-sections of the atomic arrangement in the precipitate have been plotted in Fig. \ref{slipA} along the slip plane A at different stages of the shearing of the precipitate by the dislocations. The cross-sections in Figs. \ref{slipA}(a) and (d) correspond to the initial atomic arrangement of the precipitate, while those in Figs. \ref{slipA}(b) and (e)  show the strain indicated by the blue circles in Figs. \ref{MS}(a) and (b). Finally,  Figs. \ref{slipA}(c) and (f) show the atomic arrangement of the precipitate at the end the simulations in Fig. \ref{MS} after four dislocations sheared the precipitate.  Shearing of the precipitate took place along the (110) plane and it was localized along one crystallographic plane indicated by the broken blue line in Fig. \ref{SlipPlanes}(b). The atoms moved in this plane along the [$\bar 1$10] and [00$\bar 1$] directions and the overall shear took place along the [1$\bar 1$1] direction, which is parallel to the Burgers vector of the edge dislocation in Mg (Fig. \ref{Rot}). This process was repeated as more dislocations overcame the precipitate, leading to the shearing of the precipitate. These results indicate that precipitate shearing by basal dislocations in Mg-Al alloys is favored because the (0001) Mg basal  plane is parallel to the (110) crystallographic plane of the precipitate. 

\begin{figure}[t]
	\centering
	\includegraphics[width=0.9\textwidth]{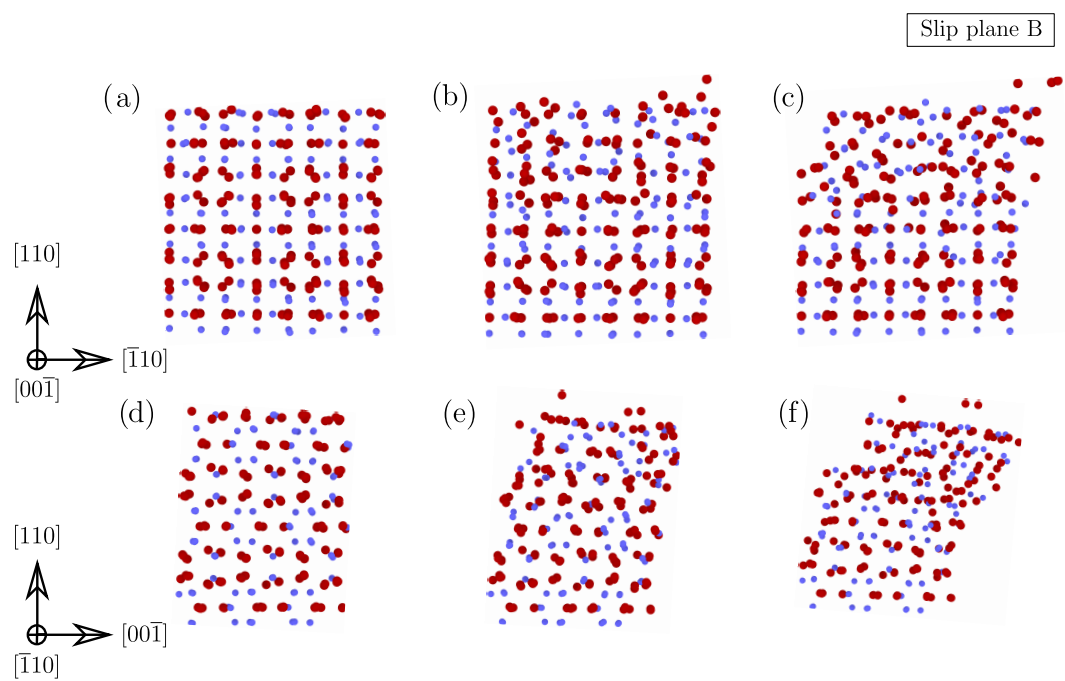}
	\caption{Shearing of the precipitate along the slip plane B. (a) and (d) Initial cross-sections.  (b) and (e)  cross- sections for the strain marked by a red circle in Figs. \ref{MS}(a) and (b)). (c) and (f) Cross sections  at the end the molecular statics simulations after four dislocations have bypassed the precipitate. The crystallographic directions in this figure correspond to the $\beta$ precipitate.}
	\label{slipB}
\end{figure}

The cross-sections of the precipitate that was sheared along the slip plane B are depicted in  Fig. \ref{slipB} at different stages of the shearing process: initial configuration in (a) and (d), at the strain marked by the red circles in Fig. \ref{MS}a) in (b) and (e) and at the end of the simulations in (c) and (f). Shear deformation also occurred parallel to (110) planes  but it was not localized in one single plane and it was difficult to assess the dominant orientation of the shear deformation. Moreover, the long range order in the upper part of the precipitate was lost after several dislocations sheared the precipitate  (Figs. \ref{slipB}(c) and (f)). Thus, the shearing mechanism of the precipitate depends on the location of the slip plane in the Mg matrix with respect to the precipitate lattice but the shear stress necessary to overcome the precipitate was similar for slip planes A and B.  Moreover the maximum shear stress (which is the  CRSS ) was attained after the second dislocation bypassed the precipitate in both cases.

\section*{References}

\end{document}